\documentclass[aps,prl,twocolumn,showpacs,showkeys]{revtex4}

\usepackage{amsmath,amssymb,epsfig,amsfonts}
\usepackage{graphicx}

\begin{document}

\title{Continuous Centrifuge Decelerator for Polar Molecules}

\author{S.~Chervenkov}
\email{sotir.chervenkov@mpq.mpg.de}
\author{X.~Wu}
\author{J.~Bayerl}
\author{A.~Rohlfes}
\author{T.~Gantner}
\author{M.~Zeppenfeld}
\author{G.~Rempe}

\affiliation{Max-Planck-Institut f\"ur Quantenoptik, Hans-Kopfermann-Str. 1, D-85748 Garching, Germany}

\begin{abstract}
Producing large samples of slow molecules from thermal-velocity ensembles is a formidable challenge. Here we employ a centrifugal force to produce a continuous molecular beam with a high flux at near-zero velocities. We demonstrate deceleration of three electrically guided molecular species, CH$_3$F, CF$_3$H, and CF$_3$CCH, with input velocities of up to $200\,\rm{m\,s^{-1}}$ to obtain beams with velocities below $15\,\rm{m\,s^{-1}}$ and intensities of several $10^9\,\rm{mm^{-2}\,s^{-1}}$. The centrifuge decelerator is easy to operate and can, in principle, slow down any guidable particle. It has the potential to become a standard technique for continuous deceleration of molecules.
\end{abstract}



\maketitle
The emerging research field of cold polar molecules promises new insights into fundamental physics and novel applications ranging from quantum simulations to controlled chemistry~\cite{Carr2009,Jin2012}. A key approach to obtaining cold molecules is the deceleration of molecular beams. This has been achieved with several techniques~\cite{Gupta1999,Bethlem1999,Fulton2004,Narevicius2008a,Hogan2009}, but only in the pulsed mode with a very low duty cycle. Applying these techniques to continuous molecular sources is highly inefficient in that only a small fraction of the available flux is utilized. To make full use of such sources, continuous deceleration is needed. This would boost current experiments~\cite{Vutha2010,Hudson2011} and open up new avenues for research, e.g., with dense ensembles accumulated in a trap~\cite{Englert2011}, in analogy to the Zeeman slower which revolutionized atomic physics 3 decades ago~\cite{Phillips1982}.

Here we present a general and versatile continuous deceleration technique for polar molecules. The molecules are decelerated by the inertial force on a rotating disk while propagating from its periphery towards its center in an electrostatic quadrupole guide with a spiral shape. An additional annular guide around the periphery of the rotating disk enables continuous operation. The capabilities and the universality of our technique are demonstrated by deceleration of three species, CH$_{3}$F, CF$_{3}$H, and CF$_{3}$CCH, from a liquid-nitrogen-cooled source~\cite{Junglen2004a} with different initial kinetic energies of the order of $100\,\rm{K}$. Output beams with intensities of several $10^9\,\rm{mm^{-2}\,s^{-1}}$ for molecules with kinetic energies below $1\,\rm{K}$ are achieved. Even higher intensities are expected for molecules from a supersonic beam or a cryogenic buffer-gas cell~\cite{Maxwell2005,vanBuuren2009}.

\begin{figure*}[ht]
\centering
\includegraphics[width=1.0\linewidth]{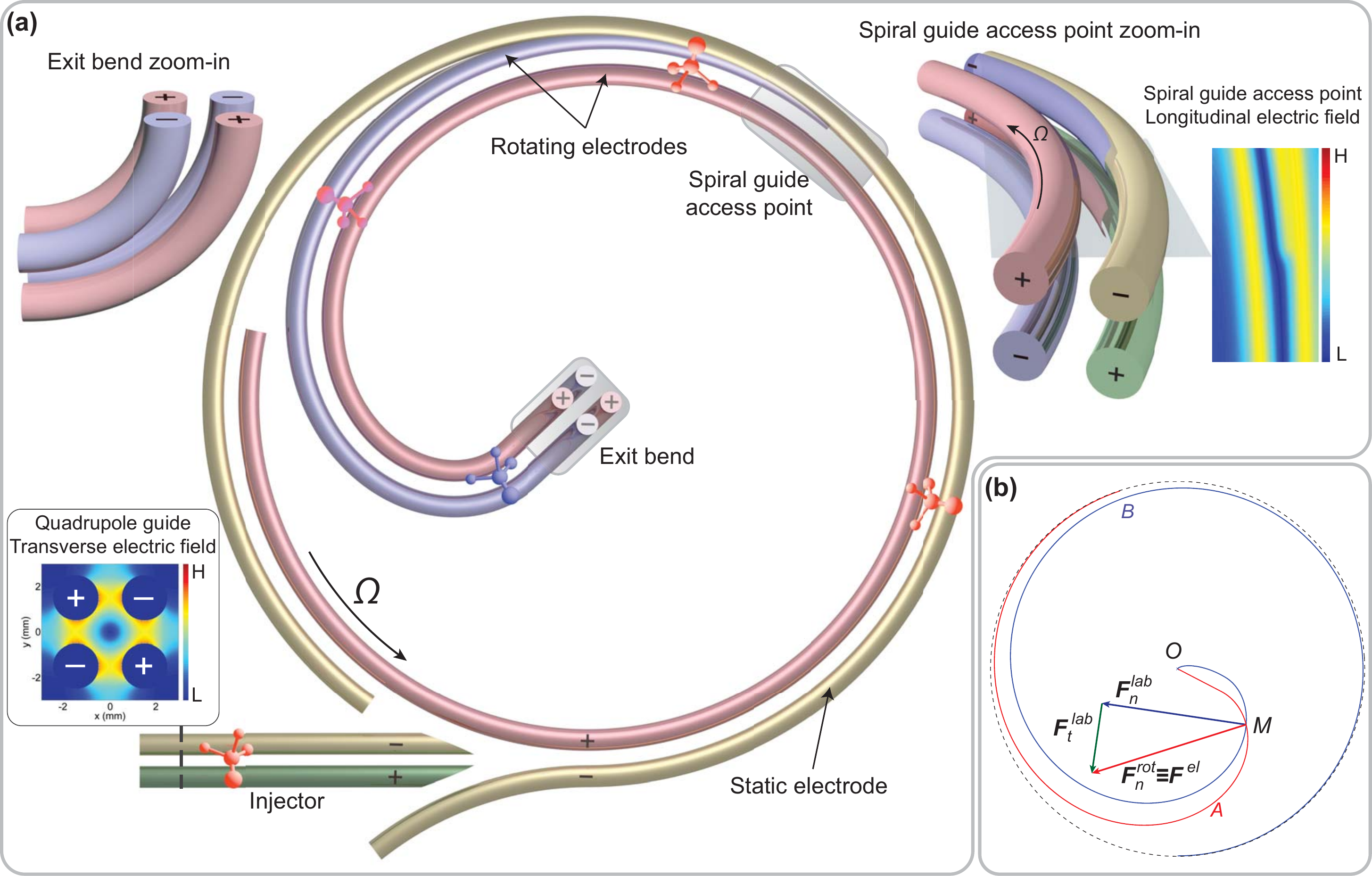}
\caption{(Color online). (a) Centrifuge decelerator, top view. For a description of the molecules' propagation through the decelerator, see text. $\Omega$ is the centrifuge rotation speed. Color bar labels {\it L} and {\it H} stand for low and high electric field, respectively. (b) Force vector diagram explaining the origin of the deceleration force (for details, see text). Red curve {\it A}: spiral-shaped quadrupole guide (``longitudinal'' molecular trajectory in the rotating frame) rotating with an angular velocity $\Omega$ about the axis $O$; blue curve {\it B}: trajectory in the laboratory frame for a given molecule's input velocity $v_{\rm{in}}$ and a centrifuge angular velocity $\Omega$.}
\label{CentrifugeSetup}
\end{figure*}

The principle of operation and the technical design of the centrifuge decelerator are schematically depicted in Fig.~\ref{CentrifugeSetup}(a). A continuous beam of polar molecules is produced by a liquid-nitrogen-cooled effusive source and velocity prefiltered~\cite{Junglen2004a} in a bent quadrupole guide (prefilter) with a radius of curvature of $20\,\rm{cm}$ (not shown in the figure). The beam is delivered to the centrifuge decelerator by a straight piece of an electric quadrupole guide dubbed injector. The transverse electric-field distribution of the injector is shown in an inset in Fig.~\ref{CentrifugeSetup}(a). The molecular beam is launched into a peripheral annular guide encompassing the centrifuge. This guide is built up of two static (outer) electrodes and two rotating (inner) electrodes [see Fig.~\ref{CentrifugeSetup}(a)]. The transverse electric-field distribution in the annular guide is the same as the one in the injector. Molecules in low-electric-field-seeking states propagate along the peripheral guide, and those of them that are fast enough catch up with the access point [see Fig.~\ref{CentrifugeSetup}(a), access point highlighted] of a rotating spiral-shaped quadrupole guide and embark onto the rotating frame. The access point is the interface between the annular and the rotating quadrupole guides, where the outer two electrodes of the rotating spiral guide approach most closely the two static electrodes. The gap between the tapered rotating electrodes and the static electrodes is $\sim200\,\rm{\mu m}$ [see Fig.~\ref{CentrifugeSetup}(a), access point zoom-in]. The small gap size ensures a seamless transfer of the molecules from the laboratory to the rotating frame due to the continuous guiding potential, as can be seen from the plot of the electric-field distribution in the plane containing the center line of the annular guide. The transfer can occur almost at any point around the periphery of the centrifuge, which makes possible the operation of the centrifuge decelerator in the continuous regime.

The deceleration of the molecules takes place in the rotating spiral quadrupole guide while they propagate from the periphery to the center of the centrifuge. As they climb up the centrifugal potential hill, they transform their initial kinetic energy into a centrifugal potential one, $E_{\rm{pot}}=-\frac{1}{2}m(\Omega r)^2$, where $m$ is the mass of the molecule, $\Omega$ is the magnitude of the centrifuge's angular velocity, and $r$ is the distance to the centrifuge's axis of rotation. The centrifuge potential is a conservative one, and hence implies conservation of the phase-space density of the decelerated molecules. We emphasize that in the rotating frame the electric field of the quadrupole guide serves only to keep the molecules in the guide and does not decelerate them, since the electric force $\boldsymbol{F}^{\rm{el}}\equiv\boldsymbol{F}_{\rm{n}}^{\rm{rot}}$ [Fig.~\ref{CentrifugeSetup}(b)] is normal to the quadrupole guide at any point and hence it does not perform work. It is solely the centrifugal potential that decelerates the molecules. In the laboratory frame, however, the electric force does the deceleration. The electric force vector $\boldsymbol{F}^{\rm{el}}$ is not normal to the longitudinal velocity of the molecules in that frame, and hence it has a component $\boldsymbol{F}_{\rm{t}}^{\rm{lab}}$ that is antiparallel to the molecules' longitudinal velocity. This is the force that performs the deceleration [Fig.~\ref{CentrifugeSetup}(b)]. Note that the angular momentum of the molecules with respect to the centrifuge's axis of rotation is decreased by the torque that $\boldsymbol{F}^{\rm{el}}\equiv\boldsymbol{F}_{\rm{n}}^{\rm{rot}}$ exerts on them. Finally, the decelerated molecules are extracted from the centrifuge along its axis of rotation. This is realized through an exit bend with a radius of curvature of $5\,\rm{cm}$ lying in a plane containing the axis of rotation [Fig.~\ref{CentrifugeSetup}(a), exit bend zoom-in], which ends up with a $1$-cm-long straight section along the rotation axis [not shown in Fig.~\ref{CentrifugeSetup}(a)]. Our centrifuge has a radius $R=0.20\,\rm{m}$ and has been operated so far at rotation speeds of up to $50\,\rm{Hz}$, well below the design speed.

\begin{figure}[t]
\centering
\includegraphics[width=1.0\linewidth]{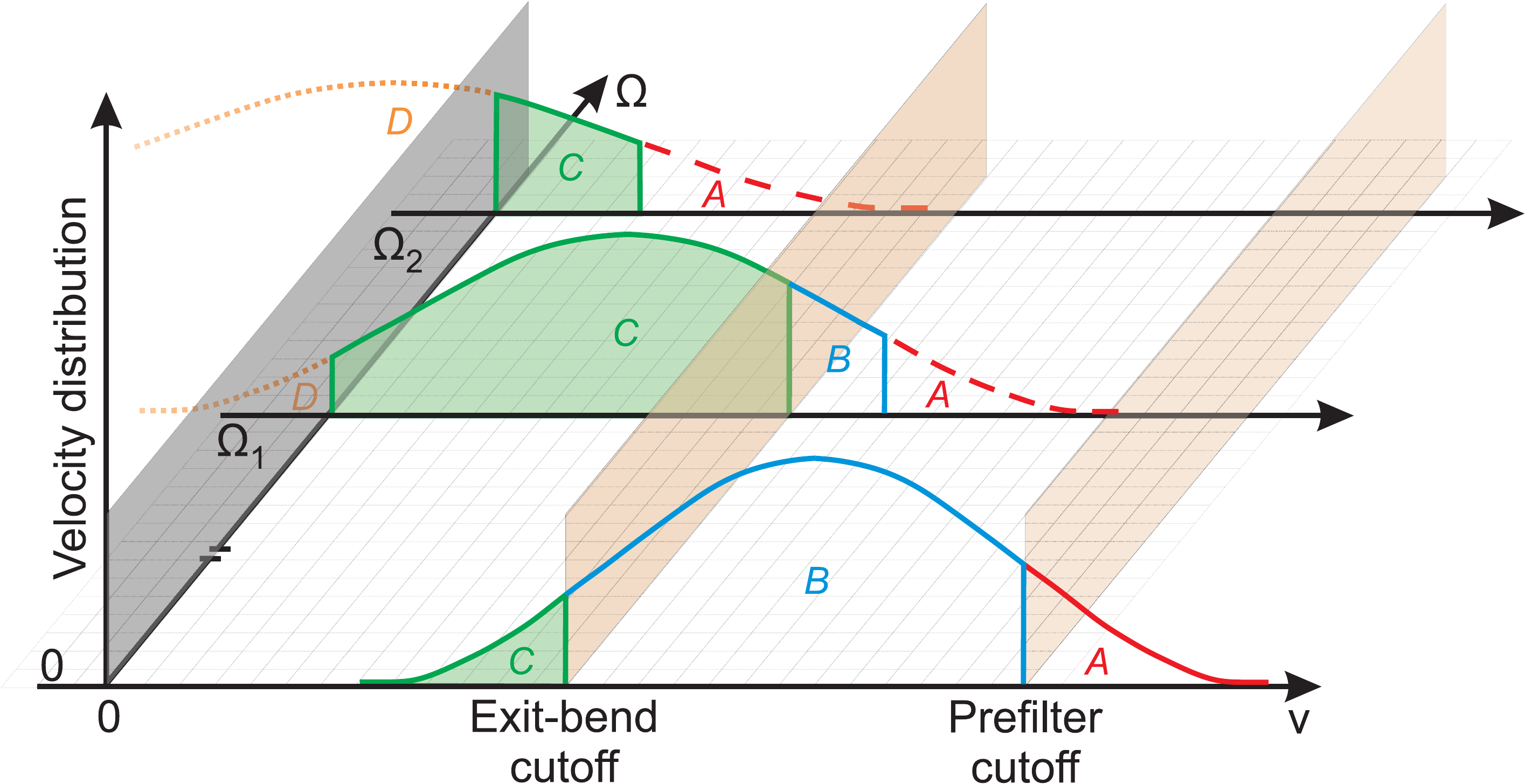}
\caption{ (Color online). Sketch of the output velocity distributions for different centrifuge rotation speeds $\Omega$. The centrifuge deceleration results in a shift to lower velocities and a broadening of the velocity distribution. Red lines {\it A} and blue lines {\it B} designate the fractions of the velocity distributions cut off by the prefilter and by the exit bend, respectively (see text). Green lines and green shaded areas {\it C} show fractions of the velocity distribution exiting from the centrifuge. Orange dotted lines {\it D} display molecules with insufficient kinetic energy, which do not reach the exit of the centrifuge, and are reflected back towards the source.}
\label{ArtistView}
\end{figure}

The following model rationalizes the experimental results from the centrifuge decelerator. Figure~\ref{ArtistView} presents schematically the centrifuge deceleration effect and the formation of the output velocity distribution for different rotation speeds of the centrifuge. For simplicity, we assume that all molecules are in a single guidable state. The molecules pass through two low-pass velocity filters on their way from the source to the exit of the centrifuge: the prefilter (with the same radius of curvature as the annular guide) before the centrifuge and the exit bend of the centrifuge [cf. Fig.~\ref{CentrifugeSetup}(a)]. The cutoff velocities for the two filters are designated in Fig.~\ref{ArtistView} as the vertical planes ``Prefilter cutoff'' and ``Exit-bend cutoff'', respectively. The spiral guide has been designed such that its cutoff velocity at any point is always between the cutoff velocities of the prefilter and the exit bend. The prefilter truncates the velocity distribution of the molecular beam injected into the centrifuge. The fraction of the initial velocity distribution cut off by the prefilter is shown by the red curves {\it A} in Fig.~\ref{ArtistView}. For a nonrotating centrifuge ($\Omega=0$) (Fig.~\ref{ArtistView}, lower curve), the molecules move through the centrifuge with unchanged velocities. The exit bend truncates the velocity distribution further, transmitting only a small fraction of it (Fig.~\ref{ArtistView}, lower curve, green shaded area {\it C}). The fraction of the velocity distribution cut off by the exit bend is designated by the blue part {\it B} of the curves in Fig.~\ref{ArtistView}. This fraction can be considered as a ``reservoir'' of fast molecules. The deceleration effect of the spinning centrifuge manifests itself in a shift to lower velocities and a broadening of the velocity distribution. The latter reflects the fact that for a constant flux a decrease of velocity must result in an increase of spatial density. For the phase-space density to be conserved, the velocity distribution must broaden. For a moderate rotation speed $\Omega_1$ (see Fig.~\ref{ArtistView}, middle curve), most of the molecules from the prefilter-truncated velocity distribution reach the exit bend with velocities smaller than its cutoff velocity. This gives rise to an increase of the output flux of slow molecules (Fig.~\ref{ArtistView}, middle curve, green shaded area {\it C}). The orange dotted line {\it D} shows the fraction of molecules with insufficient kinetic energy to reach the center of the centrifuge. Spinning the centrifuge even faster leads to a stronger deceleration, which results in a larger shift and broadening of the velocity distribution. For a rotation speed $\Omega_2$ (Fig.~\ref{ArtistView}, upper curve), even the fastest molecules from the truncated input velocity distribution are transmitted by the exit bent guide. At this rotation speed, however, most of the molecules do not have enough kinetic energy to reach the center of the rotating spiral guide (Fig.~\ref{ArtistView}, upper curve, dotted orange line {\it D}), and are reflected from the centrifugal barrier back towards the source. As a consequence, the output flux of slow molecules drops down (Fig.~\ref{ArtistView}, upper curve, green shaded area {\it C}). This effect could be interpreted also as a depletion of the reservoir of fast molecules.

\begin{figure}[t]
\centering
\includegraphics[width=0.9\linewidth]{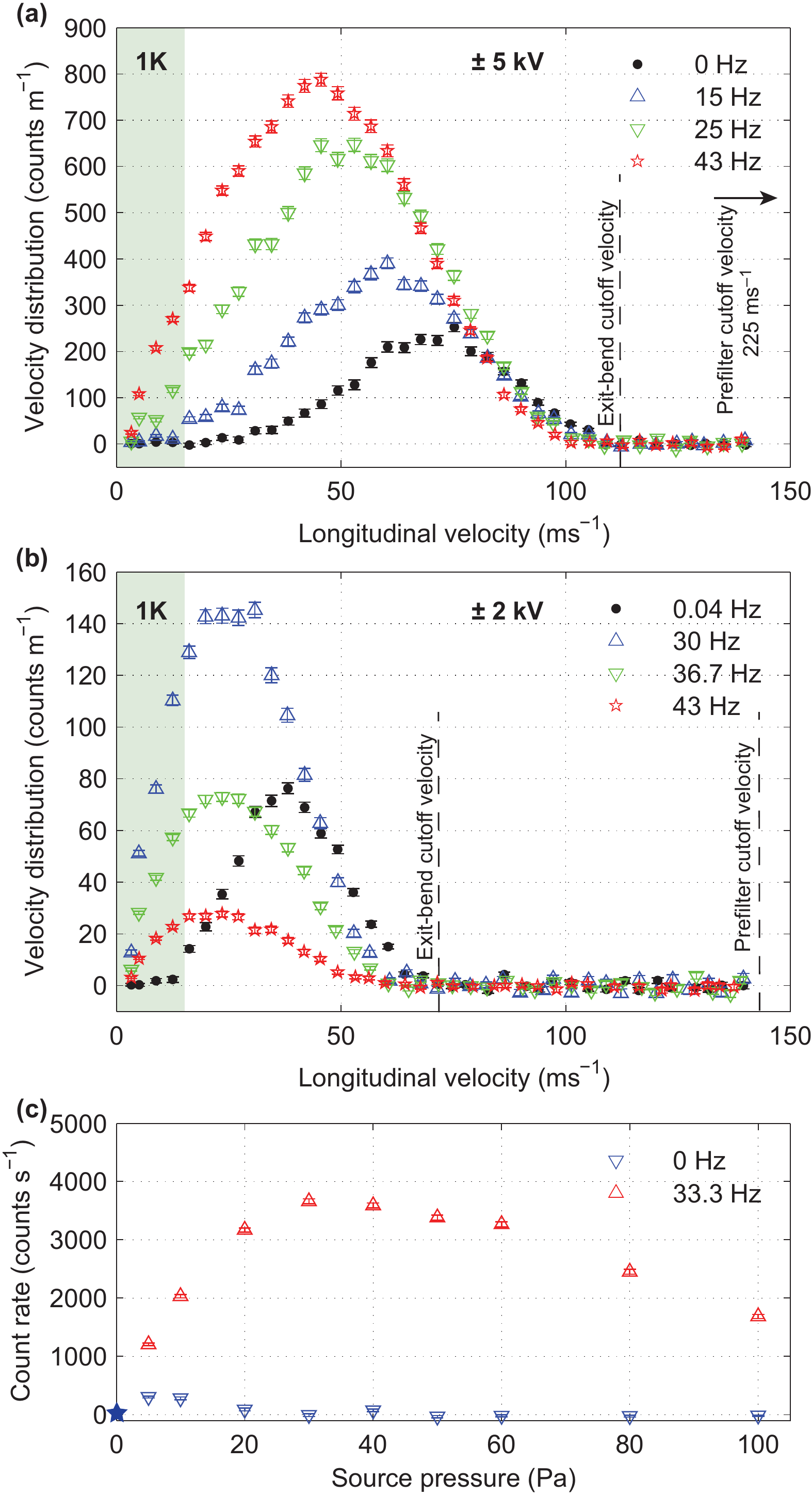}
\caption{(Color online). Experimental results for CF$_{\boldsymbol{3}}$H ($\rm{m}=70\,\rm{u}$). (a) Velocity distributions of the continuous output beam of CF$_3$H molecules at a guide voltage of $\pm 5\,\rm{kV}$ for different centrifuge rotation speeds. The range of velocities trappable in a 1-K-deep trap is highlighted in green. (b) Velocity distributions of the continuous output beam of CF$_3$H molecules at a guide voltage of $\pm 2\,\rm{kV}$ for different centrifuge rotation speeds. The range of velocities trappable in a 1-K-deep trap is highlighted in green. (c) Nozzle-pressure dependence of the integrated count rate of molecules with longitudinal velocities between 0 and $15\,\rm{m\,s^{-1}}$ (corresponding to an energy of $1\,\rm{K}$) for the centrifuge at rest and for the centrifuge at a rotation speed of $33.3\,\rm{Hz}$. The blue asterisk designates the lack of signal at zero pressure. All quoted errors stem from counting statistics of the time-binned time-of-flight signal.}
\label{Distributions}
\end{figure}

Compelling evidence for the centrifuge deceleration is provided by the output velocity distributions of CF$_3$H ($\rm{m}=70\,\rm{u}$, $d=1.65\,\rm{Dy}$) for different centrifuge rotation speeds [Fig.~\ref{Distributions}(a)] at a guide voltage of $\pm5\,\rm{kV}$ (maximum electric-field strength $\sim90\,\rm{kV\,cm^{-1}}$), corresponding to a guide depth of $1.8\,\rm{K}$ for the lowest guidable state. The molecular source was operated at a pressure $P=100\,\rm{Pa}$ (for all experiments, measured at the input of the gas line delivering the molecules to the nozzle) and a nozzle temperature $T_{\rm{noz}}=138\,\rm{K}$.  All velocity distributions in this paper were obtained by differentiation of the time-of-flight measurements in a straight quadrupole guide with a length of $50\,\rm{cm}$, immediately following the exit of the centrifuge decelerator. As a detector we used a quadrupole mass spectrometer (Hiden Analytical, HAL 301/3F) with an ionization efficiency of $\sim10^{-4}$. The experimental results excellently corroborate the model presented above (Fig.~\ref{ArtistView}). The velocity distributions correspond to the green shaded areas {\it C} in Fig.~\ref{ArtistView}. The lack of sharp falloffs at the flanks stems from the transverse spatial distribution of the molecules in the guide as well as from the distribution of the input molecules over many guidable states with different Stark shifts~\cite{MotschDepletion,Bertsche2010} and, therefore, with different cutoff velocities. The absence of slow molecules in the output signal for a nonrotating centrifuge is attributed to the boosting effect~\cite{MotschBoosting} at the molecular source. We observe very clearly a shift of the peak of the output velocity distribution and a manifold increase of the measured flux with increasing the centrifuge rotation speed. Taking into account all the factors determining the detection efficiency, we measure a total output flux of molecules with kinetic energies below $1\,\rm{K}$ (typical trap depth of electric traps~\cite{Lemeshko2013}) [highlighted in Fig.~\ref{Distributions}(a)] of the order of several $10^{9}\,\rm{s^{-1}}$ for a centrifuge rotation speed of $43\,\rm{Hz}$. From trajectory Monte Carlo simulations and from a comparison to measurements without the centrifuge, we estimate a high transmission efficiency of about $20\,\rm{\%}$. Losses occur at the transition points between the stationary and rotating guide segments and are attributed to mainly technical imperfections, which could be reduced in a next-generation centrifuge. To prove depletion of the reservoir of fast molecules at the input of the centrifuge, we have performed an experiment [Fig.~\ref{Distributions}(b)] at a guide voltage of $\pm2\,\rm{kV}$ (maximum electric-field strength $\sim36\,\rm{kV\,cm^{-1}}$), corresponding to a guide depth of $0.7\,\rm{K}$ for the lowest guidable state, leading to a decrease of the cutoff velocity and a significant truncation of the input velocity distribution. The molecular beam source was at $P=30\,\rm{Pa}$ and $T_{\rm{noz}}=138\,\rm{K}$. As seen, for rotation speeds above $30\,\rm{Hz}$ the signal progressively drops down. This result corresponds to the case presented by the upper curve in Fig.~\ref{ArtistView}.

\begin{figure}
\centering
\includegraphics[width=1.0\linewidth]{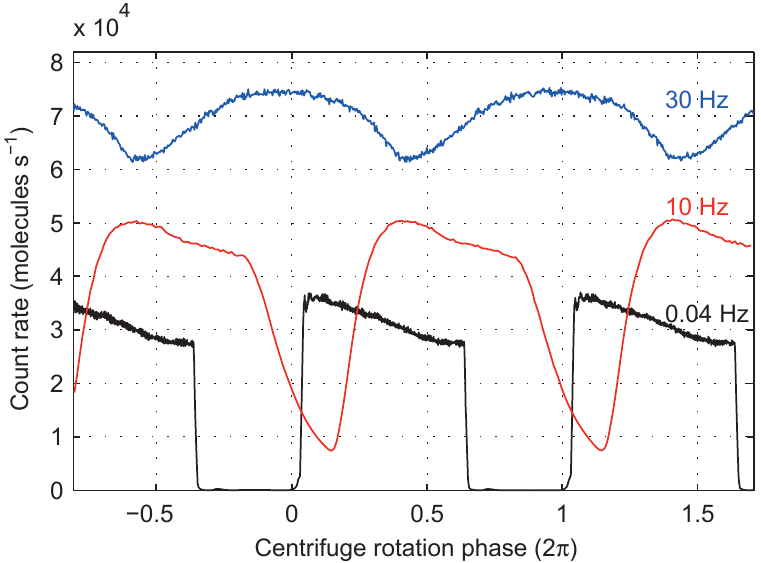}
\caption{(Color online). Dependence of the output flux of CF$_{\boldsymbol{3}}$H on the centrifuge rotation phase for different centrifuge rotation speeds. For $2/3$ of the rotation cycle the molecules experience a continuous guiding potential around the periphery of the centrifuge provided by the stationary annular guide encompassing it. With increasing rotation speed, a phase shift is observed, the total count rate rises, and the modulation depth decreases. For further explanations, see text.}
\label{FluxRotPhase}
\end{figure}

A further manifestation of the centrifuge's deceleration effect is provided by the source-pressure dependence of the flux of slow $\rm{CF}_3\rm{H}$ molecules with velocities below $15\,\rm{m\,s^{-1}}$ (kinetic energy of $1\,\rm{K}$) obtained for the centrifuge at rest and for the centrifuge spinning at $33.3\,\rm{Hz}$ under identical beam-source conditions (nozzle temperature $T_{\rm{noz}}=125\,\rm{K}$.) [Fig.~\ref{Distributions}(c)]. The signals obtained in the two cases differ dramatically in their amplitudes. The increase of the source pressure entails a shift of the velocity distribution of the source output beam to higher velocities (stronger boosting effect) thus depleting the number of slow molecules entering the centrifuge to practically zero and, consequently, decreasing the measured signal of slow molecules for $\Omega=0$ (no change in kinetic energy of the guided molecules) [cf. Figs.~\ref{Distributions}(a) and~\ref{Distributions}(b), black dots]. For the centrifuge spinning at $33.3\,\rm{Hz}$ the deceleration defeats the boosting effect, and thus many initially fast molecules injected into the centrifuge exit from it with velocities below $15\,\rm{m\,s^{-1}}$, this giving rise to a large signal increase for slow molecules in the above velocity interval.

To prove the continuous output of the centrifuge decelerator, a continuous input beam of CF$_3$H molecules has been injected into the centrifuge. Three independent measurements of the output fluxes have been carried out for three centrifuge rotation speeds of $0.04\,\rm{Hz}$, $10\,\rm{Hz}$, and $30\,\rm{Hz}$, respectively. The output fluxes as a function of the rotation phase of the centrifuge are shown in Fig.~\ref{FluxRotPhase}. The phase is defined as the angular displacement between the injector and the access point [cf. Fig.~\ref{CentrifugeSetup}(a)]. Within the range of rotation speeds, the measured signal increases with the increase of the rotation speed as a result of the centrifuge's deceleration effect. The modulation of the signal stems from the presence of sectors around the periphery of the centrifuge without a guiding potential. At high rotation speeds ($30\,\rm{Hz}$), we observe a continuous output signal with a small periodic modulation superimposed on it. The signal does not drop down to zero, because the time the molecules spend in the rotating spiral guide becomes comparable to or longer than the centrifuge's rotation period. An additional reason is also the finite width and the broadening of the velocity distribution.

To demonstrate the versatility of the centrifuge decelerator, we have proven deceleration also for two other polar species, CH$_3$F ($\rm{m}=34\,\rm{u}$, $d=1.85\,\rm{Dy}$) and CF$_3$CCH ($\rm{m}=94\,\rm{u}$, $d=2.36\,\rm{Dy}$), examples for a lighter and a heavier molecule, respectively. The velocity distributions and the count rates obtained for different rotation speeds are very similar to the ones for CF$_3$H shown in Fig.~\ref{Distributions}(a).

The centrifuge decelerator demonstrated here employs an inertial force, a universal means for deceleration of any guidable particle, be it molecules, atoms or even neutrons~\cite{Lavelle2010}. Once assembled, the decelerator does not require a sophisticated control, it is robust, and it is easy to operate. Our device has been operated for a few hundred hours to date without malfunctions. The method could, in principle, be extended by implementing rotating guide fields on a chip. We plan to combine the centrifuge decelerator with our hydrodynamically enhanced cryogenic buffer-gas source of molecules~\cite{Sommer09,Hutzler2011}. This should result in large fluxes of slow molecules with high internal-state purity. Accumulation of the molecules in an electric trap~\cite{Englert2011} and further cooling~\cite{Zeppenfeld2012} might allow us to dramatically increase the phase-space density for controlled collision experiments with polyatomic molecules and pave the way to achieving quantum degenerate regimes with polar molecules.

We thank C. Sommer, T. Wiesmeier, and M. Wismer for help at various stages of the experiment.

\bibliographystyle{unsrt}

\end{document}